\documentclass[pre,twocolumn,groupedaddress,showpacs]{revtex4}
\usepackage{graphicx,epsfig,amsmath}

\begin{document}
\bibliographystyle{apsrev}

\title{Preferred orientation of n-hexane crystallized in silicon nanochannels:\\ A combined x-ray diffraction and sorption isotherm study}

\author{Anke\,Henschel}
\author{Pushpendra\,Kumar}
\author{Tommy\,Hofmann}
\author{Klaus\,Knorr}
\email{knorr@mx.uni-saarland.de}
\author{Patrick\,Huber}
\email{p.huber@physik.uni-saarland.de}
\affiliation{Faculty of Physics and Mechatronics Engineering, Saarland University, D-66041 Saarbr\"ucken, Germany}

\date{\today}

\begin{abstract}
We present an x-ray diffraction study on $n$-hexane in tubular silicon channels of approximately 10~nm diameter both as a function of the filling fraction $f$ of the channels and as a function of temperature. Upon cooling, confined $n$-hexane crystallizes in a triclinic phase typical of the bulk crystalline state. However, the anisotropic spatial confinement leads to a preferred orientation of the confined crystallites, where the $<$001$>$ crystallographic direction coincides with the long axis of the channels. The magnitude of this preferred orientation increases with the filling fraction, which corroborates the assumption of a Bridgman-type crystallization process being responsible for the peculiar crystalline texture. This growth process predicts for a channel-like confinement an alignment of the fastest crystallization direction parallel to the long channel axis. It is expected to be increasingly effective with the length of solidifying liquid parcels and thus with increasing $f$. In fact, the fastest solidification front is expected to sweep over the full silicon nanochannel for $f=1$, in agreement with our observation of a practically perfect texture for entirely filled nanochannels.
\end{abstract}


\maketitle


Molecular ensembles embedded in pores and channels of a few nanometer diameter solidify in crystal structures which are close if not identical to those of the bulk solid state. This has been established e.g. for Ar, CO, and O$_{\rm 2}$ in porous varieties of silica with pore diameters of $7-10$\,nm \cite{Huber99,Knorr03}. C$_{\rm 16}$H$_{\rm 34}$ and other medium length $n$-alkanes embedded in Si nanochannels still show the layered crystal structures with a tight quasi-hexagonal lateral arrangement of the molecules within the layers known from the bulk state \cite{Henschel,Sirota1,Sirota2,Doucet1,Doucet2,Dirand}. Thus the primary principles of crystalline packing are still obeyed in nanochannels, similarly as has been reported for $n$-alkanes confined in microemulsions \cite{nAlkanEmulsionen} and in microcapsules \cite{nAlkanMicrocapsules}. Pore confinement introduces however a certain degree of (quenched) disorder that leads to lattice defects and lattice strains, an increased $T$-range of disordered phases (the liquid state in general, the rotator phases of the alkanes \cite{Huber2005}, the plastic phases of  N$_{\rm 2}$ and CO \cite{Huber1998,Huber99a}, and the isotropic or paranematic phase of liquid crystals \cite{Kityk2008}).\\

Quite naturally the question arises whether the crystal lattice of the solidified pore filling has a preferential orientation with respect to the pore axis. A suitable substrate to answer this question is porous Si. Electrochemical etching of (100) Si wafers using appropriate etching conditions leads to an array of parallel channels perpendicular to the surface of the wafer \cite{Lehmann,Kumar2008,Zhang2000,Cullis1997}.

$n$-alkanes solidified in such channels develop a well defined crystalline texture which we interpreted in terms of a ''nano''-version of the Bridgman process of single crystal growth \cite{Bridgman1925}. The idea is that eventually those crystallites will prevail, which are orientated in such a way that the direction of fastest crystallization is along the pore axis \cite{Henschel}. Clearly such a selection by growth is most efficient in case the solidification front can propagate over long distances. This situation was fulfilled in the study of the $n$-alkanes since here the pores had been filled completely by spontaneous imbibition, i.e. over the full length of the pores \cite{Huber2007, Henschel}. The Bridgman process is expected to be less efficient in partially filled pores. Therefore, we felt motivated to investigate the crystallization behavior in silicon nanochannels as a function of the degree of channel filling. Controlled partial filling of the channels can be achieved by condensation out of the vapor phase. With the equipment at our disposal this filling technique cannot be applied to the medium length alkanes, we rather had to switch to hexane which has a reasonably high vapor pressure down to temperatures sufficiently far below room temperature.

\begin{figure}[hbt]
\begin{center}
\includegraphics[scale=0.4]{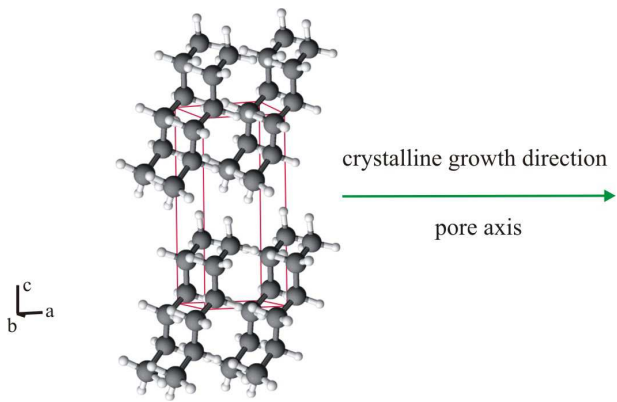}
\caption{\label{structure} \small{(Color online). Orientation of the triclinic unit cell of $n$-hexane crystallized in tubular silicon nanochannels. Note, the crystalline (100) direction of $n$-hexane, which is the direction of fastest crystal growth, coincides with the long axis of the channel.}}
\end{center}
\end{figure}

Bulk hexane (and the other even-numbered $n$-alkanes up to C$_{\rm 18}$H$_{\rm 38}$) solidify directly into a triclinic crystal structure \cite{Nyburg}. The molecules are arranged side by side in layers with a collective tilt of the molecular axis with respect to the layer normal (Fig.~\ref{structure}).

\begin{figure}[hbt]
\begin{center}
\includegraphics[scale=0.38]{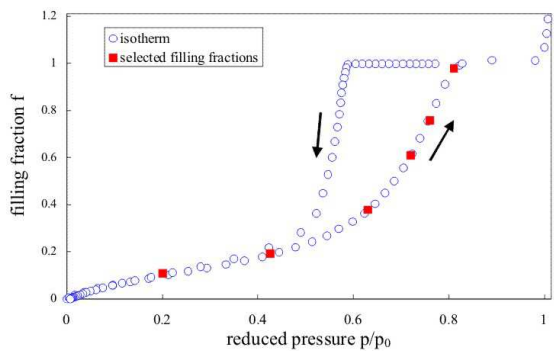}
\caption{\label{isotherm} \small{(Color online) Volumetric sorption isotherm of hexane 273~K.}}
\end{center}
\end{figure}

In Fig.~\ref{isotherm} we show a volumetric sorption isotherm of hexane in porous Si taken at 273\,K. The uptake is normalized to the value at complete filling, the vapor pressure $p$ to the saturated vapor pressure $p_{\rm 0}$ at this temperature. The initial reversible part of the isotherm is due to the growth of an adsorbed film on the pore walls, the hysteretic part to the formation of the capillary condensed fraction of the pore filling on $p$-increase and its evaporation on $p$-decrease. The analysis of the isotherm and complementary information from electron microscopy suggests an average pore diameter of 10\,nm and a porosity of 50\%. The pores extend into the wafer by 250\,$\mu$m. Hence the length-to-width ratio of the pores is more than $10^{4}$.
\begin{figure}[hbt]
\begin{center}
\includegraphics[scale=0.45]{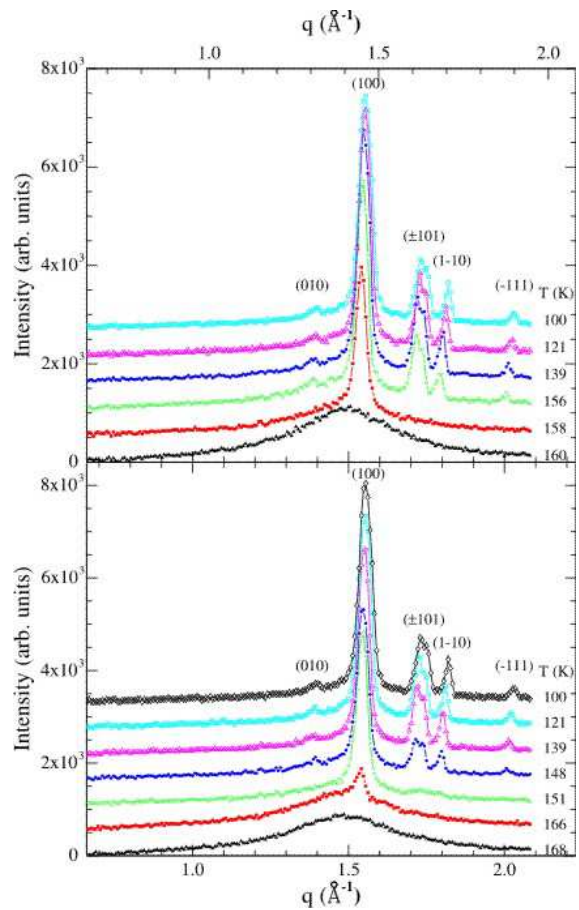}
\caption{\label{filling0.6} \small{(Color online) Diffraction patterns at selected temperatures with filling fraction 0.61 on cooling (upper panel) and heating (lower panel).}}
\end{center}
\end{figure}
Selected fractional fillings (marked in Fig.~\ref{isotherm}) have been investigated by x-ray diffractometry at temperatures ranging from 180\,K (in the liquid regime) to 100\,K (in the solid regime).  Coupled 2$\theta$-$\theta$ scans have been performed with the scattering vector oriented perpendicular to the wafer, i.e. parallel to the pore axes. Unfortunately, the diffraction setup employed did not allow for rocking scans, that is a variation in the sample/cryostat orientation at a fixed 2$\theta$. Figure~\ref{filling0.6} shows diffraction patterns at selected temperatures of the cooling/heating cycle of the sample with a fractional filling of $f=0.61$. The powder lines are indexed in terms of a triclinic ($P\overline{1}$) structure with the lattice parameters as given in \cite{Norman} ($a=4.17$\,\r{A}, $b=4.70$\,\r{A} $c=8.57$\,\r{A}, $\alpha=96.6\,^{\circ}$, $\beta=87.2\,^{\circ}$, $\gamma=105.0\,^{\circ}$).

\begin{figure}[hbt]
\begin{center}
\includegraphics[scale=0.4, angle=90]{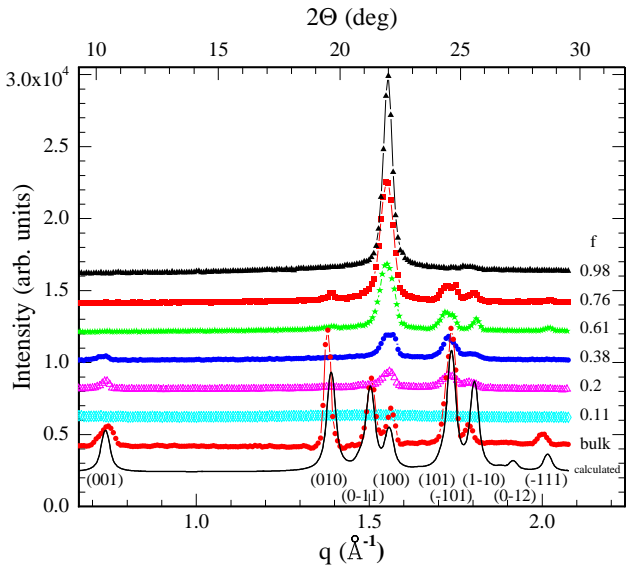}
\caption{\label{vergleich} \small{(Color online) Diffraction patterns at 100~K of all filling fractions $f$ (including the calculated and measured bulk pattern).}}
\end{center}
\end{figure}

In Fig.~\ref{vergleich} the low-$T$ patterns of all filling fractions investigated are compiled and compared to two powder diffraction patterns of the bulk state. The first one is calculated from the structural data of \cite{Norman} and the second, experimental one is obtained by means of a rapid heating/cooling cycle that triggers a distillation process upon which some of the pore solid evaporates and re-condenses on the surface of the Si wafer, presumably as "snow" \cite{Huber99}. Bragg reflections are absent for the fractional filling of 0.11 which falls into the regime of film growth (see the isotherm, Fig.~\ref{isotherm}). It is the capillary condensed fraction of the pore filling, only, that solidifies into a crystalline state.

\begin{figure}[hbt]
\begin{center}
\includegraphics[scale=0.8]{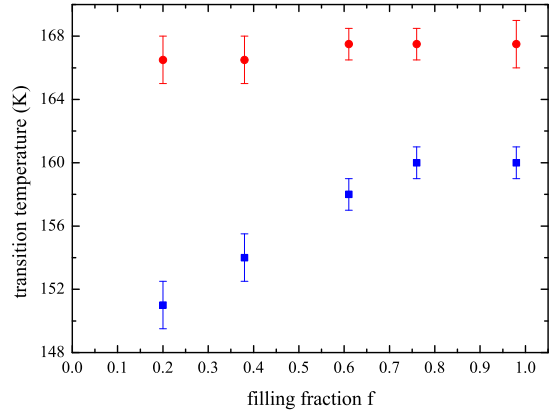}
\caption{\label{temperatures} \small{(Color online) Transition temperatures on cooling (squares)/heating (circles) of pore-confined hexane depending on the filling fraction $f$.}}
\end{center}
\end{figure}

In Fig.~\ref{temperatures} we show the freezing and the melting temperatures, $T_{f}$ and $T_{m}$, of the crystalline part as a function of $f$. $T_{f}$ decreases with decreasing $f$ and both melting and freezing occurs over a final temperature range. We simply identify $T_{f}$ ($T_{m}$) as the temperature where the first (last) indications of Bragg reflections appear (disappear) on cooling (heating). The occurrence of a final freezing range is due to the pore size distribution $P(R)$. The width of this distribution is finite, as can be already gathered from the finite slopes of the branches of the hysteresis loop of the sorption isotherm. Given such a distribution and referring to the $1/R$ scaling of the Kelvin and the Gibbs-Kelvin equation suggest the following scenario: the narrow pore segments are the first to fill on adsorption. For finite partial fillings, $0.2\leq f\leq1$, the capillary condensate consists of isolated parcels of liquid residing in those segments along the pores that have the smallest cross section available. On cooling these parcels then solidify sequentially, starting with those in the segments with the largest cross section. The length of the occupied segments increases with increasing $f$ and approaches the full pore length for $f\rightarrow$1. In the range of freezing and melting diffraction patterns have been taken in steps of 2\,K, hence the values cited have an uncertainty of this size. Bulk hexane melts at 178\,K. Thus the liquid-solid transition of the pore filling is lowered with respect to the bulk state, and there is hysteresis between freezing and melting. The same observation has been made for a larger number of pore fillings \cite{Christenson2001,AlbaSim2006,Knorr2008}.
\\

Even though this will be of no direct importance for the following, one notes that $T_{m}$ on the other hand is about independent of $f$.  The melting of pore fillings is more collective an effect than freezing, as has been, e.g., demonstrated in Refs. \cite{Schaefer} for Ar in SBA-15 by calorimetry. Cycling through the solid state may even change the connectivity of the pore filling as is suggested by optical transmission studies on Ar in Vycor glass \cite{Soprunyuk}.

Returning to the diffraction patterns one notes that whenever Bragg reflections of the pore solid could be detected, they line up with reflections of bulk hexane (Fig.~\ref{temperatures}). Hence the crystal structure of the pore confined solid is that of the bulk solid, with lattice parameters identical within experimental accuracy. The triclinic structure of hexane is obviously very robust with respect to the disorder induced by pore confinement. However, as far as Bragg intensities are concerned the bulk situation is never reproduced.  We ascribe this observation to a crystallographic texture of the pore solid, with the (100) plane being preferentially aligned perpendicular to the pore axis. This preference increases with increasing $f$ and is almost perfectly obeyed for $f=1$. Here the diffraction pattern is dominated by the (100) reflection; other reflections are practically absent. At lower $f$, additional reflections show up, but the stronger reflections are still those with a Miller index $h=\pm1$, namely, ($\bar{1}$01) and (1$\bar{1}$0), in addition to the (100) reflection which persists to be the strongest one down to $f=0.2$.\\

The preferred orientation of the pore crystals is visualized in Fig.~\ref{structure}. The direction of preferred growth lies within the layers of the layered crystal structure and coincides roughly with close-packed rows of molecules within such layers. This also means that the energy of the terminating surface in that direction is lower than the surface energy of any other crystal facet.

\begin{figure}[hbt]
\begin{center}
\includegraphics[scale=0.8]{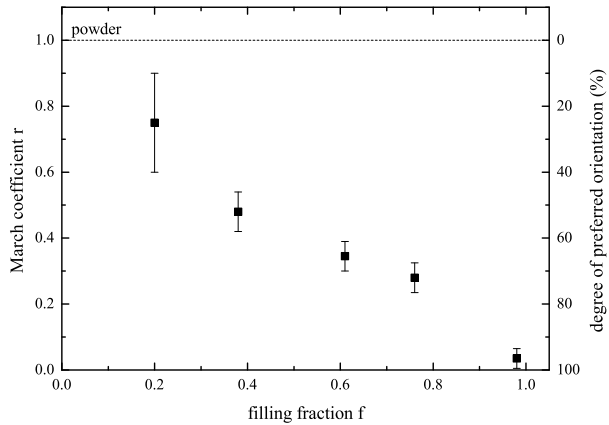}
\caption{\label{texturfilling} \small{Texture parameter as a function of filling fraction as determined from a comparison with texture powder diffractometry.}}
\end{center}
\end{figure}

It is possible to quantify the degree of this preferred orientation by resorting to models for anisotropic crystallite orientation distributions and their impact on powder diffractometry. Here, we resort to the March model \cite{MarchModel} as implemented in the program Powder Cell \cite{PowderCell}. We calculated for each filling fraction diffraction patterns and optimized the texture parameter $r$ of this model in order to achieve the best agreement between measured and calculated diffraction pattern. The texture parameter, which is 1 for a perfect powder and 0 for a perfect alignment of all crystallites with the $<$100$>$ direction along the channel long axes, is plotted in Fig \ref{texturfilling}. In agreement with the qualitative discussion above, the texture increases monotonically with $f$, and we reach a 97\% alignment for $f=1$.

Analogous observations have been made for medium length $n$-alkanes \cite{Henschel} and n-alkanols \cite{Henschel2008}. The present Brief Report supplies additional information on how the texture increases with increasing filling fraction. The results strongly support an interpretation in terms of a Bridgman-type growth process of the pore crystallites.  The propagation of the solidification front is fastest along (100)  with the result that those crystallites will finally prevail, which are oriented in such a way that this crystallographic direction is parallel to the pore axis. The length of liquid parcels increases with increasing filling fraction, hence the degree of preferred orientation increases with the filling fraction, finally leading to practically perfect texture for $f=1$ where the solidification front can sweep over the full pore length.

It should be mentioned that similar textures have been reported for the crystallization of folded polymer chains forming lamellae in aligned tubular alumina pores \cite{Steinhart}. Therefore, we believe our findings presented here for a simple $n$-alkane is of more general interest. For the future we suggest to perform crystallization studies on more complex molecules confined in porous silicon, in particular on proteins. The crystallization of protein solutions in the proximity of mesoporous silicon attracted attention only recently, for it could be demonstrated that these matrices can promote the formation of protein bulk crystallites in the surrounding bulk solutions \cite{Chayen2001}. How this effect may be related to a possible texture formation within the channels, may be an interesting question to address in the future.

\begin{acknowledgments}
The work was supported by the Deutsche Forschungsgemeinschaft via the Sonderforschungsbereich 277, Saarbr\"ucken.
\end{acknowledgments}

\end{document}